\documentclass[12pt]{article}
\usepackage{amsmath,amssymb}
\usepackage[english]{babel}
\usepackage[cp866]{inputenc}
\usepackage{hyperref}

\numberwithin{equation}{section}
\newcommand{\bq}{\begin{eqnarray}}
\newcommand{\eq}{\end{eqnarray}}
\newcommand{\bbq}{\begin{equation*}}
\newcommand{\eeq}{\end{equation*}}
\newcommand{\ra}{\rightarrow}
\newcommand{\ov}{\overline}

\newcommand{\la}{\Lambda_Q}

\newcommand{\lym}{\Lambda_{SYM}}

\newcommand{\wt}{\widetilde}

\newcommand{\ml}{m_L}
\newcommand{\mh}{m_H}

\begin{document}

\begin{center} {\bf \Large Phase transitions between confinement and higgs phases} \end{center}

\begin{center} {\bf \Large in ${\cal N}=1\,\, SU(N_c)$ SQCD with $1\leq N_F\leq N_c$ quark flavors} \end{center}

\vspace*{3mm}

\begin{center}{\bf Victor L. Chernyak} $^{\,a,\,b}$ \end{center}

\vspace*{3mm}

\begin{center} $^a\,$ Budker Institute of Nuclear Physics SB RAS, 630090 Novosibirsk, Lavrent'ev ave.11, Russia\end{center}

\begin{center} $^b\,$ Novosibirsk State University, 630090 Novosibirsk, Pirogova str.2, Russia\end{center}

\vspace*{3mm}

Talk given at the International Conference on Particle Physics and Cosmology (professor V.A. Rubakov memorial conference), October 02-07, 2023, Yerevan, Armenia.\\

This talk includes main results only. To all those who are interested in further details, all details can be found in these
two not large articles in arXiv.

\hspace*{0.5cm} {\bf Based on papers \cite{ch1}= arXiv:2109.14238 and \cite{ch2}= arXiv:2306.12994 }.\\

\begin{center}{\bf Abstract} \end{center}

Considered is the standard 4-dimensional  ${\cal N}=1$ supersymmetric $SU(N_c)$ QCD (SQCD) with $1\leq N_F\leq N_c$ quark flavors with masses $m_{Q,i}=m_{Q,i}(\mu=\la)$ in the bi-fundamental representation,  where $\la$ is the scale factor of the gauge coupling in the UV region, see e.g. \cite{ADS},\cite{S1},\cite{S2}. {\bf The gauge invariant order parameter $\rho$ is introduced in \cite{ch1} distinguishing confinement (with $\rho=0$) and higgs (with $\rho\neq 0$) phases}.

Using a number of independent arguments for different variants of transition between the confinement and higgs regimes in these theories, it is shown that {\bf transitions between these  regimes are  not crossovers but the phase transitions}.\\

In \cite{FS} the very special $SU(N_c)$ QCD theory with $N_F=N_c$  defective scalar "quarks" in the unitary gauge ~: \, $\Phi^i_{\beta}= \delta^i_{\beta}(|v|={\rm const}) > 0,\,\,i,{\beta}=1,...,N_F=N_c$, was considered by E.Fradkin and S.H.Shenker. The conclusion of \cite{FS} was that { \bf the transition between the confinement at $0 < |v|\ll\Lambda_{QCD}$ (according to \cite{FS}) and higgs (i.e. with condensed quarks) at $|v|\gg\Lambda_{QCD}$ regimes is the analytic crossover}.  And although the theory considered in [3] was very specific, the experience shows that up to now there is a widely spread opinion that this conclusion has general applicability.

This model \cite{FS} is criticized  in \cite{ch1},\cite{ch2} as incompatible with and very different from the standard  non-SUSY $SU(N_c)$ $\,\,N_F=N_c$ QCD theory with standard scalar quarks $\phi^i_{\beta}$ with all $2 N^2_c$ their physical real degrees of freedom. It is emphasized that this model \cite{FS} is really {\bf the Stuckelberg  $SU(N_c)$ YM-theory with no dynamical electric quarks and massive all $N_c^2-1$ electric gluons with fixed by hands nonzero masses}. There is no genuine confinement in this theory, it stays permanently in the completely higgsed (i.e. condensed) by hands phase only. And this is a reason for a crossover in this theory. While in the theory with standard scalar quarks there is the phase transition between the confinement (at $0 < |v|\ll\Lambda_{QCD}$) and higgs (at $|v|\gg\Lambda_{QCD}$) regimes.

Besides, the arguments presented in \cite{IS} by K.Intriligator and N.Seiberg for the standard direct  $SU(N_c)$, $\,\,N_F=N_c\,\,\, {\cal N}=1$ SQCD in support of the crossover from \cite{FS} are criticized in \cite{ch2} as erroneous.

\tableofcontents
\numberwithin{equation}{section}

\section{Definitions}

The direct $SU(N_c)$ SUSY QCD (SQCD) with $1\leq N_F\leq N_c$ quark flavors with masses $m_{Q,i}$ in the fundamental representation is defined in a standard way. The Lagrangian at the scale $\mu=\la$ looks as
\bq
K={\rm Tr}\,\Bigl (Q^\dagger e^{V } Q+ (Q\ra {\ov Q}) \Bigr )\,, \quad {\cal W}={\cal W}_{SYM}+{\rm Tr}\,({\ov Q}m_{Q}Q)\,.   \label{(1.1)}
\eq

The Konishi anomaly looks as
\bq
m_{Q,i}\langle M^i_i\rangle=\langle S\rangle,\quad M^i_j=M^i_j(\mu=\la)=\sum_{\beta=1}^{N_c}{\ov Q}^{\beta}_j Q^i_{\beta},  \label{(1.2)}
\eq
and $\langle S=\lambda\lambda/32\pi^2\rangle$ is the gluino condensate with its universal value \cite{VY},\cite{ADS},\cite{ch}
\bq
\langle S\rangle=\lym^3=\Bigl (\la^{3N_c=N_F}\Pi_{i=1}^{N_F}m_{Q,i} \Bigr )^{1/N_c}\,.  \label{(1.3)}
\eq

Besides, we use the perturbatively exact NSVZ $\beta$ function of ${\cal N}=1$ SQCD \cite{NSVZ} for the gauge coupling $a=N_c g^2/8\pi^2=N_c\alpha/2\pi$
\bq
\frac{da(\mu)}{d\ln\mu}=\beta (a)=a^2\,\frac{(3N_c-N_F)-N_F\gamma_Q(a)}{N_c(1-a)}\,, \label{(1.4)}
\eq
where $\gamma_Q$ is the quark anomalous dimension.\\

The dual non-gauge Seiberg's theory of mesons and baryons is obtained from his dual theory with $N_F=N_c+1$ flavors \cite{S1} at large mass $m_{Q,i=N_c+1}=\la$ of one extra quark flavor in the direct theory with $N_F=N_c+1$ quark flavors . The Lagrangian at the scale $\mu=\la$ looks as ($N_L+N_H=N_c,\,\, m_L\ll m_H\ll\la$)
\bq
{\wt K}_{N_c+1}= {\rm Tr}_{N_c+1}\frac{M^\dagger M}{\la^2}+\sum_{i=1}^{N_F=N_c+1}\Bigl ( (B^\dagger)^i B_i +({\ov B}^\dagger)_i\,{\ov B}^{\,i}
\Bigr ),  \label{(1.2)}
\eq
\bbq
{\wt {\cal W}}_{N_c+1}=m_L\sum_{i=1}^{N_L}  (M_L)^i_i + m_H\sum_{i=1}^{N_H}  (M_H)^i_i + \la M^{N_c+1}_{N_c+1}+{\rm Tr}_{N_c+1}\,({\ov B}\frac{M}{\la} B)-\frac{\det_{N_c+1} M}{\la^{2N_c-1}}\,.
\eeq

\section{Direct theory with $1\leq N_F\leq N_c-1$ quark flavors}

In this range of $N_F$, the weak coupling higgs phase at $\mu_{\rm gl}\gg\la$ for light equal mass quarks with $0 < m_Q\ll\la$ looks as follows. All quarks are higgsed, i.e. form a constant coherent condensate in a vacuum state in the logarithmic weak coupling regime. And the perturbative pole masses of $N_F(2N_c-N_F)$ massive gluons look as (with logarithmic accuracy, $\langle S\rangle$ is the standard gluino condensate, $M^i_j=\sum_{\beta=1}^{N_c}({\ov Q}^{\beta}_{j}Q^i_{\beta})$
\bq
\Bigl (\frac{\mu_{\rm gl}}{\la}\Bigr )^2 \sim g^2(\mu=\mu_{\rm gl})\,z_Q(\la,\mu=\mu_{\rm gl})\, \frac{\rho_{\rm higgs}^2} {\la^2} 
\sim\frac{1}{N_c}\frac{\langle M\rangle}{\la^2}\sim \label{(2.1)}
\eq
\bbq
\sim\frac{1}{N_c}\frac{\langle S\rangle}{m_Q\la^2}\sim \frac{1}{N_c}\Bigl (\frac{\la}{m_Q} \Bigr )^{\frac{N_c-N_F}{N_c}}\gg 1\,. 
\eeq
Dealing with higgsed quarks, to obtain \eqref{(2.1)}, we first separate out Goldstone fields from quark fields $Q^i_{\beta}$ normalized at the scale $\la$ (a part of these Goldstone fields or all of them will be eaten by gluons  when quarks are higgsed)
\bq
Q^i_{\alpha}(x)=\Bigl (G_{\rm Goldst}^{SU(N_c)}(x)\Bigr )^{\beta}_{\alpha}\,{\hat Q}^i_{\beta}(x)\,,\quad {\hat Q}^i_{\beta}(x)=\Bigl (
G_{\rm Goldst}^{SU(N_c)}(x)^\dagger \Bigr )^\gamma_\beta Q^i_{\gamma}(x)\,, \label{(2.2)}
\eq
\bbq
{\hat Q}^i_{\beta}(x)=\Bigl (  U_{\rm global}^{SU(N_c)}\Bigr )^{\delta}_{\beta} \Bigl ( U_{\rm global}^{SU(N_F)}\Bigr )^i_j {\tilde Q}^j_{\delta}(x),\quad \alpha, \beta, \gamma, \delta=1...N_c,\,\, i, j=1...N_F,
\eeq
where $G_{\rm Goldst}^{SU(N_c)}(x)$ is the $N_c\times N_c$ unitary $SU(N_c)$ matrix of Goldstone fields.

And then, with the standard choice of vacuum of spontaneously broken global symmetry, we replace ${\hat Q}^i_{\beta}(x)$ in \eqref{(2.2)}, containing remained degrees of freedom, by its mean vacuum value (at $\mu=\la$)
\bq
\langle {\hat Q}^i_{\beta}(x)\rangle =\langle {\hat Q}^i_{\beta}(0)\rangle=\delta^i_{\beta}\,\rho_{\rm higgs}\,, \quad \frac{\rho_{\rm higgs}}{\la}=\Bigl (\frac{\la}{m_Q}\,\Bigr )^{\frac{N_c-N_F}{2 N_c}}\gg 1\,, \label{(2.3)}
\eq
\bbq
i=1...N_F\,,\quad \beta=1...N_c\,. 
\eeq

And similarly $\langle {\hat {\ov Q}_i^{\,\beta}(x)}\rangle=\delta_i^{\,\beta}\,\rho_{\rm higgs}$.\\

Under pure gauge transformations, see \eqref{(2.2)}\,:
\bq
Q^i_\alpha (x)\ra  \Bigl ( V_{\rm pure\, gauge}^{SU(N_c)}(x)\Bigr )^\beta_\alpha Q^i_\beta (x),\quad \quad
G_{\rm Goldst}^{SU(N_c)}(x)\ra V_{\rm pure\, gauge}^{SU(N_c)}(x) G_{\rm Goldst}^{SU(N_c)}(x). \label{(2.4)}
\eq

That is, these are $Q^i_\alpha (x)$ and Goldstone fields $G_{\rm Goldst}^{SU(N_c)}$ which are transformed in the fundamental representations under pure gauge transformations,  while ${\hat Q}^i_{\beta}(x)$ stays intact under pure gauge transformations and is {\bf the colored gauge invariant quark field}. And $\rho$ in \eqref{(2.3)} is {\bf the gauge invariant order parameter}. It behaves non-analytically with varying $m_Q/\la$. It is $\rho_{\rm higgs}\neq 0$  for higgsed (i.e. condensed) scalar quarks, while $\rho_{\rm HQ}=0$ if quarks are in the HQ (heavy quark) phase and not higgsed. This non-analytic behavior is a clear sign of the phase transition.

The gauge invariant  pole masses of massive gluons are as in \eqref{(2.1)}.\\

The  gauge invariant order parameter $\rho_{\rm higgs}\neq 0$ in \eqref{(2.3)} is the counter-example to a widely spread opinion that the gauge invariant order parameter for higgsed scalar quarks in the fundamental representation does not exist.\\

It is seen from \eqref{(2.1)},\eqref{(2.3)} that  at $\mu\gg\la$ the value of the running gluon mass $\mu_{\rm gl}(\mu)\gg\la$ decreases with increasing $N_c$ and fixed $(m_Q/\la)\ll 1$.  And at sufficiently large number of colors,
\bq
N_c/(N_c-N_F)\ln (N_c) \gg \ln(\la/m_Q)\gg 1\,, \label{(2.5)}
\eq
$\mu_{\rm gl}(\mu\sim\la)$ will be much smaller than $\la$. This means that even quarks with large $(\rho_{\rm higgs}/\la)=\Bigl (\la/m_Q\,\Bigr )^{(N_c-N_F)/2 N_c}\gg 1$ are not higgsed then in the weak coupling regime at $\mu\gg\la$. And now, at such $N_c$ \eqref{(2.5)}, all quarks and gluons will remain effectively massless at scales $\lym\ll m^{\rm pole}_Q < \mu < \la$. 

Let us recall a similar situation at $N_c < N_F < 3N_c/2$ considered in section 7 of \cite{ch} (only pages 18 - 21 including the footnote 18 in arXiv:0712.3167\, [hep-th]). As pointed out therein, when decreasing scale $\mu$ crosses $\mu\sim\la$ from above, the increasing perturbative coupling $a(\mu)$ crosses unity from below. But for (effectively) massless quarks and gluons the perturbatively exact NSVZ $\beta$-function \cite{NSVZ} \eqref{(1.4)} can't change its sign by itself (and can't become frozen at zero outside the conformal window) and behaves smoothly. I.e., when increased $a(\mu)$ crosses unity from below and denominator in \eqref{(1.4)} crosses zero, increased quark anomalous dimension $\gamma_Q(\mu)$ crosses $(3N_c-N_F)/N_F$ from below, so that the $\beta$-function behaves smoothly and remains negative at $\mu < \la$.  The coupling $a(\mu\ll\la)$ continues to increase with decreasing $\mu$
\bq
\frac{d a(\mu)}{d\ln \mu} = \beta(a)\ra \, -\, \nu\, a\,<\, 0,\quad  \nu=\Bigl [\frac{N_F}{N_c}(1+\gamma^{\rm str}_Q)-3\Bigr ]=
{\rm const} > 0\,, \label{(2.6)}
\eq
\bbq
a(\mu\ll\la)\sim\Bigl (\frac{\la}{\mu} \Bigr )^{\nu\, >\, 0}\gg 1\,. 
\eeq

In section 7 of \cite{ch} (see also \cite{ch3}) the values $\gamma^{\rm str}_Q=(2N_c-N_F)/(N_F-N_c) > 1,\,\,\nu=(3N_c-2N_F)/(N_F-N_c) > 0$ at $\mu\ll\la$ and $N_c < N_F < 3N_c/2$ have been found from matching of definite two point correlators in the direct $SU(N_c)$ theory and in $SU(N_F-N_c)$ Seiberg's dual \cite{S2}. In our case here with $1 \leq N_F < N_c$ the dual theory does not exist. So that, unfortunately, we can't find the concrete value $\gamma^{\rm str}_Q$. But, as will be shown below, for our purposes it will be sufficient to have the only condition $\nu > 0$ in \eqref{(2.6)}.

As shown in \cite{ch1},  all quarks are now in the HQ (heavy quark) phase at so large $N_c$ \eqref{(2.5)}, i.e. not higgsed but confined and decouple as heavy at $\mu <  m^{\rm pole}_Q=m_Q/ z_Q(\la,m^{\rm pole}_Q\ll\la)$, where, see \eqref{(2.6)}, $\gamma_Q >\, 2$ and  $z_Q(\la,m^{\rm pole}_Q\ll\la)=(m^{\rm pole}_Q/\la)^{\gamma_Q >\, 2}\ll 1$ is the perturbative quark renormalization factor.. There remains at lower energies $\mu < m^{\rm pole}_Q\ll\la\,\,\,SU(N_c)$ SYM with $\lym^3=\langle S\rangle=\Bigl (\la^{3N_c-N_F}\Pi_{i=1}^{N_F} m_{Q,i}\Bigr )^{1/N_c}\ll\la^3$ at small $m_{Q,i}$.

The order parameter $\rho$ \eqref{(2.3)} is nonzero in the higgs phase at $\mu_{\rm gl}\gg\la$ \eqref{(2.1)}, $\rho_{\rm higgs}=
\la(\frac{\la}{m_Q})^{\frac{N_c-N_F}{2N_c}}\gg\la$, while it is zero, $\rho_{HQ}=0$, in the HQ (heavy quark) phase at fixed $m_Q/\la$ and large $N_c$, \eqref{(2.5)} (or at $m_Q\gg\la$), see sect.4.1 in \cite{ch1}. 

I.e., {\bf there is the phase transition between the phase of higgsed quarks with $\mu_{\rm gl}\gg\la$ \eqref{(2.1)} and HQ-phase with not higgsed but confined quarks at large $N_c$ \eqref{(2.5)}, or at $m_Q\gg\la$ }.\\

This phase transition is most clearly seen at $N_F=N_c-1$, see section 4.2 in \cite{ch1}.

{\bf A)}.\,\, Heavy quarks with $m_Q\gg\la$ or light quarks with fixed $m_Q\ll\la$ at large $N_c$ \eqref{(2.5)} are in the HQ-phase with $\rho_{HQ}=0$ and all confined, with the order parameter $\rho_{HQ}=0$. The global $SU(N_F)$ is unbroken. There is in the spectrum a number of heavy flavored quarkonia with typical masses ${\cal O}(m^{\rm pole}_Q)\gg\lym$ and different quantum numbers. For instance, the quark-antiquark bound states with different spins and other quantum numbers are in the adjoint or singlet representations of unbroken global $SU(N_F)$. {\bf It is important that, due to a confinement, there are no particles in the spectrum in the $SU(N_F)$ (anti)fundamental representation of dimensionality $N_F$}. Besides, there are in the spectrum a number of $SU(N_F)$ singlet gluonia  with small typical masses $\sim\lym=\la (m_Q/\la)^{N_F/3N_c}\ll\la$.

{\bf B)}.\,\, At sufficiently small $m_Q\ll\la$ and not too large $N_c$, the whole $SU(N_c)$ is broken by  higgsed quarks with $\rho_{\rm higgs}=
\la(\la/m_Q)^{1/N_c}\gg\la$. All $N_c^2-1$ gluons and their scalar superpartners acquire large masses $g\rho_{\rm higgs}\gg\la$. {\bf There
is no confinement}. They form 2 adjoint representations of $SU(N_F)$ plus two $SU(N_F)$ singlets. Plus, and this is most important, else $2 N_F$ heavy gluons $(A_{\mu})^i_{\alpha=N_c},\, (A_{\mu})^{\alpha=N_c}_{i},\, i=1...N_F$ and $2 N_F$ their ${\cal N}=1$ scalar superpartners. These $4 N_f$ form {\bf two fundamental and two antifundamental representations of $SU(N_F)$}. And finally, there are $N^2_F$ light complex pions $\Pi^i_j,\, i,j=1...N_F$ with small masses $\sim m_Q$ which form the adjoint and singlet representations of $SU(N_F)$. Therefore, there are only fixed numbers of particles with fixed quantum numbers in the spectrum. Besides, the masses of gluons in different representations of unbroken global $SU(N_F)$ are different, see sect.4.2 in \cite{ch1}.

From comparison of mass spectra in regions $m_Q\gg\la$ and $m_Q\ll\la, \mu_{\rm gl} \gg \la$ \eqref{(2.1)} it is seen that, although the unbroken global symmetry $SU(N_F)$ is the same, but realized are its different representations. In the case of heavy confined quarks in the HQ-phase there are no particles in the spectrum in the (anti)fundamental representation of $SU(N_F)$, while in the case of light higgsed quarks such representations are present. In other words. The fraction $R_{\rm fund}$ of particles in the (anti)fundamental representation of unbroken global $SU(N_F)$ in the mass spectrum can serve in the case considered as the order parameter. This fraction is zero in the confinement region where quarks with the order parameter $\rho_{HQ}=0$  are not higgsed. While this fraction is the nonzero constant in the region with higgsed quarks with $\rho_{higgs}\gg\la$. I.e., {\bf like $\rho$,  $R_{\rm fund}$ behaves non-analytically. This non-analytic behavior of the order parameter $R_{\rm fund}$ of the unbroken global $SU(N_F)$ symmetry is a clear sign of the phase transition, because this fraction would behave analytically in the case of crossover}.

\section {Mass spectra of direct and Seiberg's dual theories at $N_F=N_c$}

It is shown in \cite{ch2} that, considered as two independent theories, the low energy mass spectra at $\mu < \la$ of direct $N_F=N_c$ SQCD and Seiberg's dual theory  of $N_F^2-1$ mesons $M^i_j$ and two baryons $B,\,{\ov B}$,  are {\bf parametrically different}, both for quarks with equal or unequal masses. Therefore, {\bf the proposal by N. Seiberg \cite{S1} of this dual theory as the low energy form at $\mu < \la$ of the direct theory is erroneous}.\\

Besides, as emphasized in \cite{ch4}, the direct $SU(N_c)$ SQCD with $N_F=N_c+1$ equal mass light quark flavors and proposed by N. Seiberg in \cite{S2} as its low energy form at scales $\mu <\la$ the dual theory of $N_F^2$ mesons $M^i_j$ and $2N_F$ baryons $B_i,\,{\ov B}^j$,
$i,j=1...N_F$, have parametrically different mass spectra at $\mu <\la$. Therefore, this dual theory is not the low energy form of the direct one.\\

\section {\hspace*{-3mm} Comparison with paper of E. Fradkin and S.H. Shenker ~\cite{FS}}

In the paper \cite{FS} of E. Fradkin and S.H. Shenker, the special (non-SUSY) QCD-type lattice $SU(N_c)$ gauge theory with $N_F=N_c$ flavors of scalar quarks $\Phi^i_{\beta}$ in the bi-fundamental representation of $SU(N_c)\times SU(N_F)$ was considered. In the unitary gauge, all remained $N_c^2+1$ physical real degrees of freedom of these quarks were {\bf deleted  by hands} and replaced by one constant parameter $|v| > 0\,:\,  \Phi^i_{\beta} = \delta ^i_{\beta}|v|,\,\,\beta=1,...,N_c,\,\, i=1,...,N_F=N_c$.\, I.e., all such "quarks" are massless, with no self-interactions  and {\it permanently higgsed by hands} even at small $g|v|\ll\Lambda_{QCD}$, see page 3694 and eq.(4.1) for the bare perturbative Lagrangian in \cite{FS}.  And all $N^2_c-1$ electric gluons received  {\bf fixed by hands masses} $g |v|$. The region with $g|v|\gg\Lambda_{QCD}$ was considered in \cite{FS} as the higgs regime, while those with $0 < g|v|\ll\Lambda_{QCD}$ as the confinement one. The conclusion of \cite{FS} was that the transition between the higgs and confinement regimes is the analytic crossover, not the non-analytic phase transition. And although the theory considered in \cite{FS} was very specific, the experience shows that up to now there is a widely spread  opinion that this conclusion has general applicability.

Let us note that this model \cite{FS} looks unphysical and is incompatible with normal models with dynamical electrically charged scalar quarks $\phi^i_{\beta}$ with all $2 N_F N_c$ their real physical degrees of freedom. This model \cite{FS} is really {\bf the Stuckelberg pure $SU(N_c)$ YM-theory with no dynamical electric quarks and with massive all $N_c^2-1$ electric gluons with fixed by hands masses} $g|v| > 0$ in the bare perturbative Lagrangian, see eq.(4.1) in \cite{FS}.

For this reason, {\bf in any case}, the electric flux emanating from the test (anti)quark becomes exponentially suppressed at distances $L > l_0=(g|v|)^{-1}$ from the source. And so, {\bf the tension of the potentially possible confining string will be also exponentially suppressed} 
at distances $L > l_0$ from sources. And e.g. external heavy test quark-antiquark pair will be not connected then by one common really confining string at large distance between them. These quark and antiquark can move then practically independently of each other and can be registered alone in two different detectors at large distance between one another. I.e., {\bf in any case}, in this Stuckelberg theory \cite{FS}, at all fixed $|v| > 0$, there is no genuine confinement which prevents appearance of one (anti)quark in the far detector. Besides, the additional arguments are presented in \cite{ch2} that, because all electric gluons have non-dynamical but fixed by hands nonzero masses, then this at all prevents to form confining strings in this Stuckelberg theory.  And this is a reason for a crossover in this theory. While in the theory with standard scalar quarks there is the phase transition between the confinement and higgs regimes.\\

In support of the conclusion of E.Fradkin and S.H.Shenker \cite{FS} about the crossover between the confinement and higgs regimes, it was written by K. Intriligator and N. Seiberg in \cite{IS} for the standard ${\cal N}=1\,SU(N_c)$ SQCD with $N_F=N_c$ the following (for the moduli space, with $m_{Q,i}\ra 0)$.

''For large expectation values of the fields (the values of $\langle M^i_i\rangle$ are implied) a Higgs description is most natural while, for small expectation values, it is more natural to interpret the theory as 'confining'... Because these theories (i.e. ${\cal N}=1$ SUSY QCD) have matter fields in the fundamental representation of the gauge group, as mentioned in the introduction, there is no invariant distinction between the Higgs and the confining phases \cite{FS}. It is possible to smoothly interpolate from one interpretation to the other''.\\

In other words. Because one can move completely smoothly (i.e. analytically) on the moduli space with  e.g. $m_{L,H}\ra 0$ at fixed $r=m_L/m_H$, from the region {\bf A} with $r\ll 1,\,\,\langle M_L\rangle\gg\la^2,\, \langle M_H\rangle\ll\la^2$, see section 2.2 in \cite{ch2},
\bq
\langle M_L\rangle=\frac{\langle S\rangle_{N_c}}{\ml}=\la^2\Bigl ( \frac{1}{r}\Bigr )^{\frac{N_H}{N_c}},\,\,\,
\langle M_H\rangle=\frac{\langle S\rangle_{N_c}}{\mh}=\la^2\,\Bigl (r\Bigr )^{\frac{N_L}{N_c}}\,,
\eq
\bbq
\langle M_L\rangle\gg\la^2\,, \quad \langle M_H\rangle\ll\la^2\,, \quad r\ll 1\,, \label{(4.1)}
\eeq
where LL-quarks $Q^L_L, {\ov Q}^L_L$ are higgsed while (according to \cite{IS}) HH-quarks $Q^H_H,\, {\ov Q}^H_H$ are confined, to another region {\bf B} with $r\gg 1,\,\,\langle M_L\rangle\ll\la^2,\, \langle M_H\rangle\gg\la^2$ where, see \eqref{(4.1)}, vice versa, quarks $Q^H_H, {\ov Q}^H_H$ are higgsed while quarks $Q^L_L,\, {\ov Q}^L_L$  are confined, this is the independent confirmation of the conclusions of paper \cite{FS} that there is the analytic crossover between the confinement and higgs regimes, not the non-analytic phase transition.\\

There are two loopholes in these arguments. It is right that quarks $Q^L_L, {\ov Q}^L_L$ or $Q^H_H, {\ov Q}^H_H$ are higgsed on the moduli space in regions {\bf A} or {\bf B} with $\langle M_L\rangle\gg\la^2$ or $\langle M_H\rangle\gg\la^2$ respectively. But first, the quarks with $\langle M_{L,H}\rangle\ll\la^2$ are not confined. There are regions on the moduli space where some quarks are higgsed, but there are no regions where some quarks are confined. {\bf There is no confinement on the moduli space of the direct theory} because the tension of confining string is $\sigma^{1/2}\sim\lym\ra 0$ at $m_{Q,i}\ra 0$.\\

It was emphasized in \cite{ch} that {\bf the confinement originates only from (S)YM}. There is no confinement in Yukawa-type theories without gauge interactions. But the lower energy SYM theory contains only one universal dimensional parameter $\lym=\Bigl(\la^{3N_c-N_F}
\Pi_{i=1}^{N_F}m_{Q,i}\Bigr )^{1/3 N_c}\ll\la$ at $m_{Q,i}\ll\la$. So that, the tension of its string  can not be as large as $\sigma^{1/2}\sim\la$ at $m_{Q,i}\ll\la$, but is much smaller:\, $\sigma^{1/2}\sim\lym\ll\la$. And $\lym\ra 0$ if even one $m_{Q,i}\ra 0$, in which limit there is no confinement at all. Therefore, unlike the statements of K.Intriligator and N.Seiberg in \cite{IS}, the transitions are between regimes of higgsed or not higgsed some quarks, but in all regions of moduli space all quarks are not confined.\\

And finally, as shown in \cite{ch2}, on the way along the moduli space from region  {\bf A} to  {\bf B} at fixed $N_c$, there are {\bf two phase transitions} at the points $r=r_L(N_c)\ll 1$ and  $r=r_H(N_c)\gg 1$, where, respectively, the LL-quarks become unhiggsed and HH-quarks become higgsed. And there is the finite width region $r_L(N_c) < r < r_H(N_c)$ where all quarks are not higgsed. The gauge invariant order parameter $\rho$ \cite{ch1} is nonzero outside the region  $r_L(N_c) < r < r_H(N_c),\,\,\rho_{\rm higgs}\neq 0$, while it is zero inside it, $\rho_{HQ}=0$.\\

So that, the arguments of K.Intriligator and N.Seiberg for the  standard ${\cal N}=1\,SU(N_c)$ SQCD with $N_F=N_c$ \cite{IS}, put forward  in support of the crossover from the paper \cite{FS} of E.Fradkin and S.H.Shenker, about an analytical movement along the moduli space and an absence of phase transitions  are erroneous.

\section{Conclusions}

\hspace*{4mm}  1) The gauge invariant order parameter $\rho$ was introduced in \cite{ch1}. It is $\rho_{\rm higgs}\neq 0$ for higgsed (i.e. condensed) quarks while $\rho_{HQ}=0$ if quarks are not higgsed, independently of whether they are confined or not. This  gauge invariant order parameter $\rho_{\rm higgs}\neq 0$ is the counter-example to a widely spread opinion that the gauge invariant order parameter for higgsed scalar quarks in the fundamental representation does not exist.

2) It was shown \cite{ch1} for the standard ${\cal N}=1\,\,\, SU(N_c)$ SQCD with $1\leq N_F \leq N_c-1$ light quark flavors that there is the phase transition between the regions of higgsed quarks with $\rho_{\rm higgs}\gg\la$ at not too large $N_c$, and confined quarks in the HQ (heavy quark) phase with $\rho_{HQ}=0$ at sufficiently large $N_c$ \eqref{(2.5)} (or at $m_Q\gg\la$). And  that there is the especially clear phase transition at $N_F=N_c-1$ and fixed $N_c$ between the regions of higgsed sufficiently light quarks and confined heavy quarks.

3) It was shown \cite{ch2} for the standard ${\cal N}=1\,\,\, SU(N_c)$ SQCD with $N_F= N_c$ light quark flavors that the proposal by N. Seiberg \cite{S1} of his dual theory with only $N^2_c-1$ mesons $M^i_j$ and two baryons $B,\, {\ov B}$ as the low energy form at $\mu < \la$ of the direct $N_F=N_c$ theory is erroneous. The mass spectra of the direct $N_F=N_c$ theory and Seiberg's dual \cite{S1} are parametrically different, both for quarks of  equal or unequal masses.

4)  It was shown in detail in \cite{ch2} (see also \cite{ch1}) that considered by E.Fradkin and S.H.Shenker $N_F= N_c$ QCD theory with defective higgsed by hands scalar "quarks" is incompatible with the standard theory with quarks with all their degrees of freedom. It is really the Stuckelberg pure YM-theory with no dynamical electric quarks and with massive all $N_c^2-1$ electric gluons with fixed by hands nonzero masses.  And this is a reason for a crossover in this theory. While in the theory with standard scalar quarks there is the phase transition between the confinement and higgs regimes. 

5) It was shown \cite{ch2} that arguments presented by K.Intriligator and N.Seiberg in \cite{IS} for the standard direct  $SU(N_c)$, $N_F=N_c\,\,\, {\cal N}=1$ SQCD in support of the crossover from \cite{FS} are erroneous.

\end{document}